\documentclass[12pt]{iopart}

\usepackage{graphicx}
\usepackage{iopams}

\begin{document}

\title{A Numerical Study of Transport and Shot Noise at 2D Hopping}

\author{Yusuf A. Kinkhabwala\dag, Viktor A. Sverdlov\dag, Alexander N. Korotkov\ddag\ and Konstantin K. Likharev\dag}
\address{\dag\ Department of Physics and Astronomy, Stony Brook
University, Stony Brook, NY 11794-3800}
\address{\ddag\ Department of Electrical Engineering, University of
California - Riverside, Riverside, CA 92521}

\begin{abstract}
We have used modern supercomputer facilities to carry out extensive
Monte Carlo simulations of 2D hopping (at negligible Coulomb
interaction) in conductors with the completely random distribution
of localized sites in both space and energy, within a broad range of
the applied electric field $E$ and temperature $T$, both within and
beyond the variable-range hopping region. The calculated properties
include not only dc current and statistics of localized site
occupation and hop lengths, but also the current fluctuation
spectrum. Within the calculation accuracy, the model does not
exhibit $1/f$ noise, so that the low-frequency noise at low
temperatures may be characterized by the Fano factor $F$. For
sufficiently large samples, $F$ scales with conductor length $L$ as
$(L_c/L)^{\alpha}$, where $\alpha=0.76\pm 0.08 < 1$, and parameter
$L_c$ is interpreted as the average percolation cluster length. At
relatively low $E$, the electric field dependence of parameter $L_c$
is compatible with the law $L_c\propto E^{-0.911}$ which follows
from directed percolation theory arguments.
\end{abstract}

\pacs{72.20.Ee, 72.20.Ht, 73.50.Td}

\maketitle

\section{\label{sec:level1}Introduction}

The theory of hopping transport in disordered conductors
\cite{MottBook,ShklovskiiBook,EfrosPollackCollection} at negligible
Coulomb interaction is often perceived a well-established (if not
completed) field, with recent research focused mostly on Coulomb
effects. However, only relatively recently it was recognized that
shot noise (see, e.g., Ref.~\cite{BlanterButtiker2000}) is a very
important characteristic of electron transport. In particular, the
suppression of the current fluctuation density $S_I(f)$ at low
frequencies, relative to its Schottky formula value $2e\left\langle
I\right\rangle$ (where $\left\langle I\right\rangle$ is the dc
current), is a necessary condition \cite{MatsuokaLikharev1998} for
the so-called quasi-continuous (``sub-electron") charge transfer in
such finite-current experiments as single-electron oscillations
\cite{Av-Likh}.

Earlier calculations of shot noise at hopping through very short
samples (e.g., across thin films
\cite{1SiteNazarovStruben1996,2SiteKinkhabwalaKorotkov2000}) and
simple lattice models of long conductors in 1D
\cite{1DKorotkovLikharev2000} and 2D
\cite{2DSverdlovKorotkovLikharev2001} have shown that such
suppression may, indeed, take place. However, calculations for the
more realistic case of disordered 2D conductors have been limited to
just one particular value of electric field
\cite{2DSverdlovKorotkovLikharev2001}. It seemed important to
examine whether the law governing this suppression is really as
general as it seemed. Such examination, carried out in this work,
has become practical only after the development of a new, advanced
method of spectral density calculation
\cite{Method-KinkhabwalaSverdlovKorotkov2005} in combination with
the use of modern supercomputers. (The work reported below took
close to a million processor-hours of CPU time.)

As a useful by-product of this effort, we have obtained accurate
quantitative characterization of not only the dependence of the
average current on both temperature $T$ and electric field $E$, but
also the statistics of localized site occupation and hop lengths,
which give a useful additional insight into the physics of hopping
transport.

\section{\label{sec:level1}Model}

\begin{figure}
\begin{center}
\includegraphics[height=3.4in]{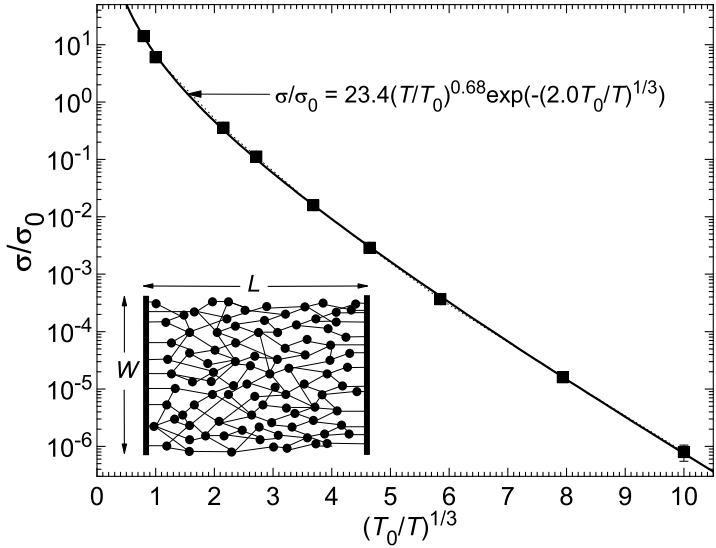}
\end{center}
\caption{Linear conductivity $\protect\sigma$ as a function of
temperature $T$. Points show the results of averaging over $80$
samples varying in size ($L \times W$) from $(20 \times 10) a^2$ to
$(160 \times 100) a^2$, biased by a low electric field $E\ll E_T$.
Points are Monte Carlo results. The thin dashed line is just a guide
for the eye, while the thick solid line corresponds to the best fit
of the data by Eq.~(\ref {eq:hightemperatureconductivity}). Here and
below, the error bars are smaller than the point size, unless they
are shown explicitly. (The error bars correspond to the uncertainty
in averaging over an ensemble of independent samples that is larger
than the calculational uncertainty for each of the samples.) The
inset shows the system under analysis (schematically).}
\label{fig:highTconductivity}
\end{figure}

We have studied hopping in 2D rectangular ($L\times W$) samples with
``open" boundary conditions on the interface with well-conducting
electrodes
 \cite{2DSverdlovKorotkovLikharev2001} - see inset in
Fig.~\ref{fig:highTconductivity}. In the present study, we have
concentrated on broad samples with width $W\gg L_c$, where $L_c$ is
the effective percolation cluster size (see below). The conductor is
assumed to be ``fully frustrated": the localized sites are randomly
distributed over the sample area, and the corresponding electron
eigenenergies $\varepsilon^{\left( 0 \right)}_j$ are randomly
distributed over a sufficiently broad energy band, so that the 2D
density of states $\nu_0$ is constant at all energies relevant for
conduction. Electrons can hop from any site $j$ to any other site
$k$ with the rate
\begin{equation}
\gamma_{jk}=\Gamma_{jk}\exp \left(-\frac{r_{jk}}{a}\right),
\label{eq:distancerateequation}
\end{equation}
where $r_{jk}\equiv \sqrt{\left( x_j -x_k\right)^2+\left( y_j -y_k
\right)^2}$ is the site separation distance, and $a$ is the
localization radius \cite{loclengthfootnote}. Such exponential
dependence on the hop length has been assumed in virtually all
theoretical studies of hopping. (The corrections to this law due to
phase interference effects \cite{Nguen85,Medina89,Wang90} are
typically small.) However, in contrast to most other authors, we
take Eq.~(\ref{eq:distancerateequation}) literally even at small
distances $r_{jk}\sim a$; this range is important only at very high
fields and/or temperatures where the average value of $r_{jk}$
becomes comparable to $a$. Of course our quantitative results for
this particular region are only true for the localized states with
exponential wavefunction decay.

Another distinction from some other works in this field is that we
assume that the hopping rate amplitude $\Gamma_{jk}$ depends
continuously on the localized site energy difference $\Delta
U_{jk}\equiv \varepsilon^{\left( 0 \right)}_{j}-\varepsilon^{\left(
0 \right)}_{k}+e\mathbf{E}\mathbf{r}_{\mathbf{jk}}$:
\begin{equation}
\hbar\Gamma_{jk}\left( \Delta U_{jk} \right)=g\frac{\Delta
U_{jk}}{1-\exp\left( -\Delta U_{jk}/k_{B}T\right) }.
\label{eq:energyrateequation}
\end{equation}
This model coincides, for low phonon energies $\Delta U_{jk}$, with
that described by Eqs. (4.2.17)-(4.2.19) of
Ref.~\cite{ShklovskiiBook} for hopping in lightly-doped
semiconductors, and of course satisfies the Gibbs detailed balance
requirement $\Gamma_{jk}=\Gamma_{kj}\exp\left( \Delta U_{jk}/k_{B}T
\right)$. It is also close, but more physical than the ``Metropolis"
dependence which has a cusp at $\Delta U_{jk}=0$.

The interaction of hopping electrons is assumed negligible (with the
exception of their implicit on-site interaction which forbids
hopping into already occupied localized states). This assumption is
well-justified for practically important materials, in particular
very thin films of amorphous silicon, which is the major candidate
material for the implementation of sub-electron charge transfer
components in single-electronic circuits \cite{LikharevIEEE1999}.
Indeed, the relative strength of the Coulomb interaction may be
characterized by a dimensionless parameter $\chi \equiv (e^2 /\kappa
a) \times (\nu_0 a^2)$, where $\kappa$ is the relative dielectric
constant \cite{KinkhabwalaSverdlovLikharev2004}. For a film of
thickness $t \sim a$, $\nu_0$ may be estimated as $t\nu$, where
$\nu$ is the 3D density of states. For undoped amorphous Si, $\nu$
is of the order of 10$^{16}$ eV$^{-1}$cm$^{-3}$, and only special
treatments may increase it to $\sim$10$^{20}$ eV$^{-1} $cm$^{-3}$ -
see, e.g., Ref.~\cite{Sameshita1991}. As a result, for $\kappa \sim
10$ and $a \sim 3$nm (both numbers are typical for the midgap states
in Si), $\chi$ is much less than unity for the entire range of $\nu$
cited above, so that there is a broad range ($\Delta \left( \ln{T}
\right), \Delta \left( \ln{E} \right) \sim 3 \ln{\left(
\chi^{-1}\right)}$) of temperature $T$ and electric field $E$ where
the Coulomb interaction is negligible
\cite{KinkhabwalaSverdlovLikharev2004}.

With this assumption, our model has only three energy scales:
$k_{B}T$, $eEa$, and $\left( \nu_0 a^{2} \right)^{-1}$. In other
words, there are two characteristic values of electric field:
\begin{equation}
E_T\equiv \frac{k_{B}T}{ea} \ \textrm{and}\ E_0\equiv
\frac{1}{e\nu_0 a^3}. \label{eq:characteristicfields}
\end{equation}
We will be mostly interested in the case of low temperatures
$T<T_0$, where
\begin{equation}
T_0\equiv \frac{1}{k_B\nu_0 a^2}
\label{eq:characteristictemperature}
\end{equation}
is the field-independent scale of temperature, so that the field
scales are related as $E_T<E_0$. (The only role of the dimensionless
parameter $g$ introduced by Eq.~(\ref{eq:energyrateequation}) is to
give the scale of hopping conductivity $\sigma_0\equiv
g(e^2/\hbar)$. Coherent quantum effects leading to weak localization
and metal-to-insulator transition are negligibly small, and hence
the formulated hopping model is adequate, only if $g \ll 1$.)

The dynamic Monte Carlo calculations were carried out using the
algorithm suggested by Bakhvalov \textit{et
al.}~\cite{Bakhvalovetal1989}, which has become the de facto
standard for the simulation of incoherent single-electron tunneling
\cite{Wasshuber}. All calculated variables were averaged over the
sample, and in most cases, over several (many) samples with
independent random distributions of localized sites in space and
energy, but with the same dimensionless parameters $L/a$, $W/a$,
$T/T_0$, and $E/E_0$. We have used a new, advanced technique
\cite{Method-KinkhabwalaSverdlovKorotkov2005} of the noise (current
spectral density) calculation to save simulation time. The used
supercomputer facilities are listed in the Acknowledgments section
below.

\section{\label{sec:level1}DC Current}

In order to understand the relation between our model and the prior
results in this field, we have started from the calculation of dc
current $\left\langle I\right\rangle$ as a function of $T$ and $E$.
If the electric field is sufficiently small ($E\ll E_T$), then the
current is proportional to $E$, and the transport is completely
characterized by the linear conductivity $\sigma\equiv \left\langle
I\right\rangle/WE$. Fig.~\ref{fig:highTconductivity} shows the
calculated conductivity as a function of temperature $T$. In the
region $T\ll T_0$ this dependence follows the exponential $T$
dependence of the 2D Mott law
\cite{MottBook,ShklovskiiBook,EfrosPollackCollection}
\begin{equation}
\frac{\sigma}{\sigma_0} \approx A\left(T,0 \right)\exp \left[-\left(
B\left(T,0 \right) \frac{T_0}{T} \right)^{1/3}\right],
\label{eq:hightemperatureconductivity}
\end{equation}
where $A\left(T, E \right)$ and $B\left(T, E \right)$ are
dimensionless, model-dependent slow functions of their arguments. We
have found that our results may be well fit by
Eq.~(\ref{eq:hightemperatureconductivity}) with the following
pre-exponential function: $A\left(T, 0 \right) = \left(23.4 \pm 1.4
\right)\left(T/T_0\right)^{\left( 0.68 \pm 0.04\right)}$, and
constant $B\left(T,0 \right)=2.0 \pm 0.2$ \cite{accuracy}. This
latter result may be compared with the following values reported in
the literature. In Ref. \cite{Brenigetal1973}, $B$ was analytically
estimated to be close to 2.1. A different value, $3.45 \pm 0.2$ has
been found by mapping a random 2D hopping problem to the problem of
percolation in a system of linked spheres \cite{SkalShklovskii1974,
ShklovskiiBook}. Finally, a close value $3.25$ (with no uncertainty
reported) has been obtained using numerical simulations of hopping
on a periodic lattice, with a slightly different model for the
function $\Gamma (\Delta U)$ \cite{TsigankovEfros2002}. The
difference between our result and the two last values is probably
due to the differences between details of the used models.

\begin{figure}
\begin{center}
\includegraphics[height=3.4in]{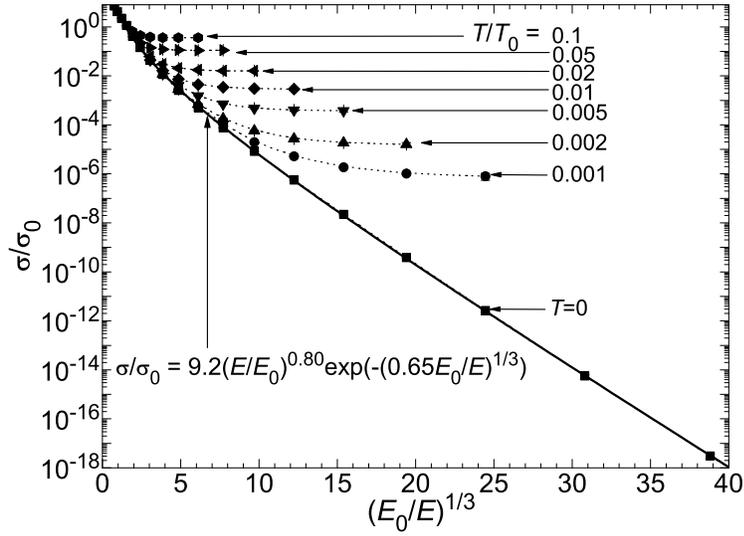}
\end{center}
\caption{Nonlinear conductivity $\sigma\equiv \left\langle
I\right\rangle/WE$ as a function of electric field $E$ for several
values of temperature $T$. Each point represents data averaged over
$80$ samples of the same size, ranging from $(20 \times 14) a^2$ to
$(1000 \times 700) a^2$, depending on $T$ and $E$. Points are Monte
Carlo results. Thin dashed lines are only guides for the eye, while
the thick solid line shows the best fit of the data by
Eq.~(\ref{eq:strongfieldconductivity}).}
\label{fig:allTandEconductivity}
\end{figure}

At higher electric fields ($E\gtrsim E_T$), dc current starts to
grow faster than the Ohm law, so that if we still keep the above
definition of conductivity $\sigma$, it starts to grow with $E$
(Fig.~\ref{fig:allTandEconductivity}). At $T\rightarrow 0$, the
results are well-described by the expression \cite{Shklovskii1973,
ApsleyHughes19741975, PollackRiess1976, RentzschShlimakBerger1979,
vanderMeerSchuchardtKeiper1982}
\begin{equation}
\frac{\sigma}{\sigma_0} \approx A \left( 0, E \right) \exp
\left[-\left( B\left(0,E \right) \frac{E_0}{E} \right)^{1/3}
\right]. \label{eq:strongfieldconductivity}
\end{equation}
The data for not very high fields $\left( E_T \ll E \ll E_0 \right)$
may be well fit by Eq.~(\ref{eq:strongfieldconductivity}) with
constant $B\left(0,E \right) = 0.65\pm 0.02$ and pre-exponential
function $A \left( 0, E\right) = \left(9.2 \pm 0.6
\right)\left(E/E_0\right)^{\left( 0.80 \pm 0.02\right)}$. (Note that
the value $B\left(0,E \right) = 1.27$ given in
Ref.~\cite{2DSverdlovKorotkovLikharev2001} corresponds to a
different pre-exponential function used for fitting.)

Finally, Fig.~\ref{fig:allTandEconductivity} shows that when the
electric field becomes comparable with the value $E_0$ defined by
the second of Eqs.~(\ref{eq:characteristicfields}), dc current and
hence, conductivity, start to grow even faster than the exponential
$E$ dependence of Eq.~(\ref{eq:strongfieldconductivity}).

\begin{figure}
\begin{center}
\includegraphics[height=3.4in]{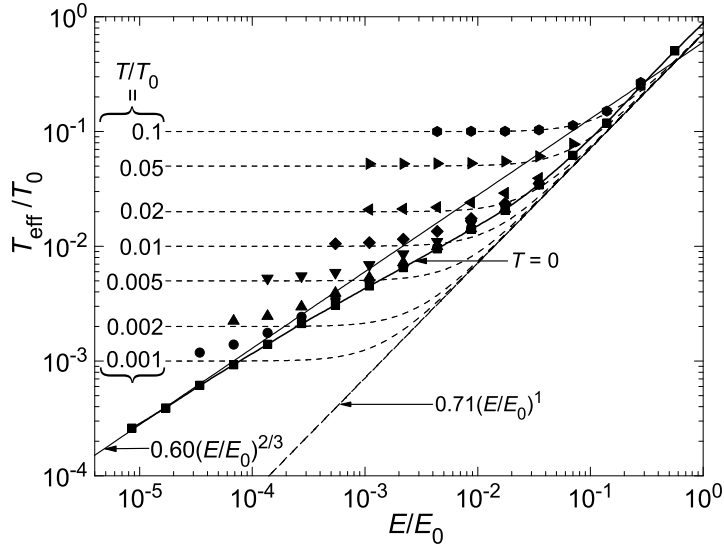}
\end{center}
\caption{The effective temperature $T_{\mathrm{eff}}$ of current
carriers as a function of electric field $E$ for several values of
temperature $T$. Closed points: Monte Carlo simulation results;
dashed lines: master equation results. The solid curve marked $T=0$
is only a guide for the eye.} \label{fig:effectivetemperature}
\end{figure}

\section{\label{sec:level1}Hopping Statistics}

\begin{figure*}
\begin{center}
\includegraphics[height=5.2in]{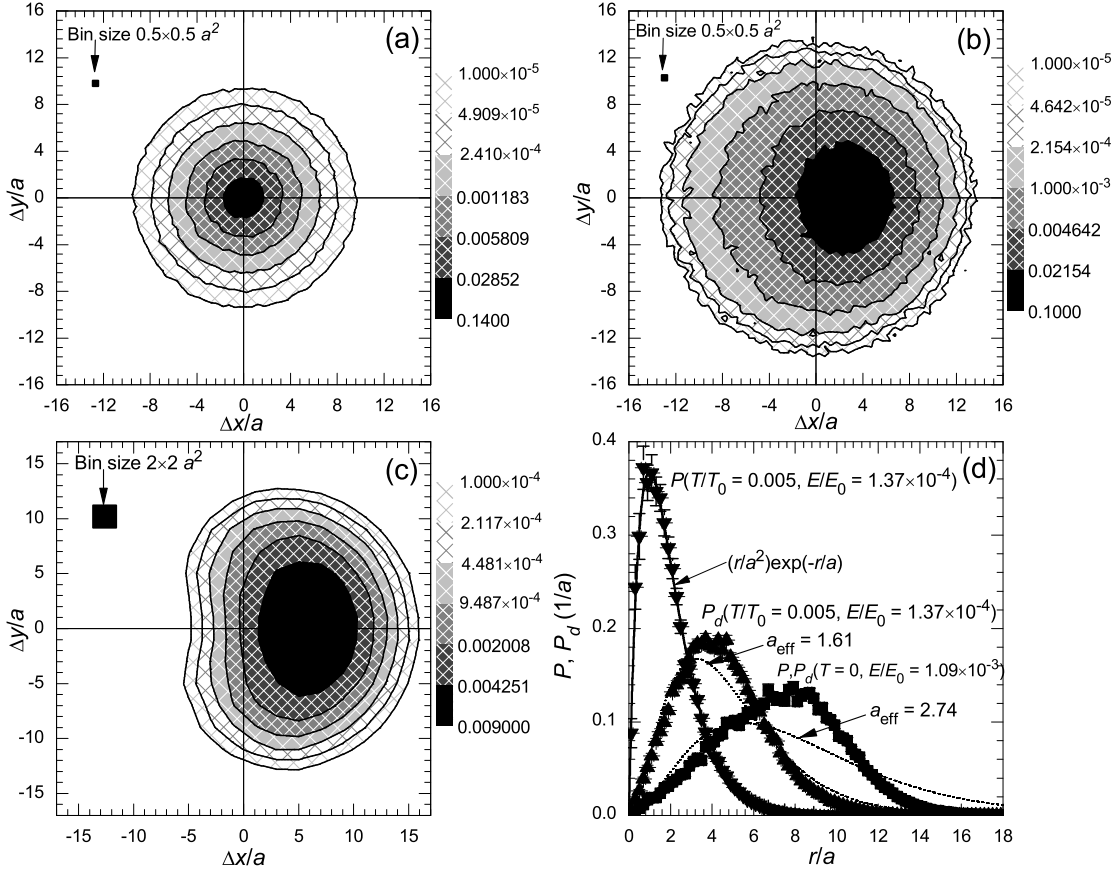}
\end{center}
\caption{(a)-(c) Two-dimensional and (d) one-dimensional histograms
of hop lengths for two typical cases: (a) and (b):
$T/T_0=5\times10^{-3}$, $ E/E_0=1.37\times10^{-4}$ ($E\ll E_T$) and
(c): $T=0$, $E/E_0=1.09 \times10^{-3}$ ($E_T\ll E$). The
shade-coding in panels (a) and (c) corresponds to the probability
$P$ of hops with given $\Delta \mathbf{r}=(\Delta x,\Delta y)$,
while that in panel (b), to the probability $P_d$ weighed by factor
$\left\vert H_{jk}-H_{kj}\right\vert$ - see the text. (Since at
$T=0$ there are no backward hops, for the case shown in panel (c),
$P$ and $P_d$ coincide.) Panel (d) shows $P$ and $P_d$, averaged
over all directions of vector $\Delta \mathbf{r}$, for the
low-field, intermediate, and high-field cases. Dashed lines show the
distribution (\ref{eq:rmdsprobdens}) given by master equation, for
the best-fit values of parameter $a_{\mathrm{eff}}$.}
\label{fig:2dand1dhophistograms}
\end{figure*}

\begin{figure}
\begin{center}
\includegraphics[height=5.4in]{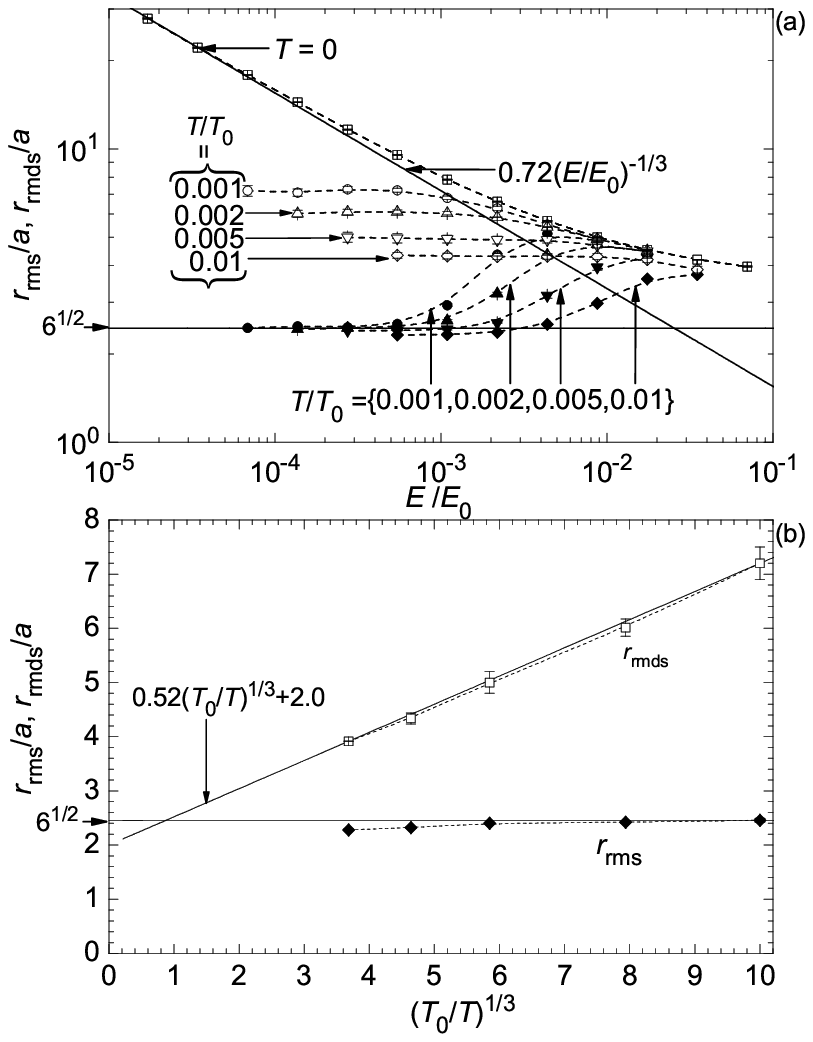}
\end{center}
\caption{R.M.S. hop length $r_{\mathrm{rms}}$ (solid points) and the
weighed average hop length $r_{\mathrm{rmds}}$ (open points) as
functions: (a) of applied electric field $E $ for several
temperatures $T$, and (b) of temperature at $E\rightarrow 0$. Tilted
straight lines in panels (a) and (b) show the best fits by
Eqs.~(\ref{eq:rrmsrrmdsstrongfield}) and~(\ref{eq:highTrrmdsfit}),
respectively, while the horizontal thin lines show the values
following from the master equation. Curves are only guides for the
eye.} \label{fig:EandThoplength}
\end{figure}

In order to understand the physics of hopping in the three field
regions better, it is useful to have a look at the statistics of
localized site occupation and hopping length. We have found that for
all studied values of $E$ and $T$, the probability of site
occupation closely follows the Fermi distribution with the local
Fermi level
\begin{equation}
\mu(\mathbf{r})=\mu_L - e\mathbf{E}\mathbf{r}
\end{equation}
(where $\mu_L$ is the Fermi level of the source electrode) and some
effective temperature $T_{\mathrm{eff}}$. Points in
Fig.~\ref{fig:effectivetemperature} show $T_{\mathrm{eff}}$ as a
function of electric field $E$ for several values of physical
temperature $T$. Dashed lines show the result of the best fitting of
the naive single-particle master equation
\begin{eqnarray}
\frac{ \partial f\left( \epsilon ,t\right) }{\partial t} = && \int
d^{2}r\int d\epsilon ^{\prime }\exp \left( \frac{-r}{a}\right)
\left\{-\Gamma \left( \epsilon -\epsilon ^{\prime }+e\mathbf{Er}
\right) f\left( \epsilon \right) \left[ 1-f\left( \epsilon ^{\prime
}\right)
\right] \right.  \nonumber \\
&& \left. \,\,\,\,\,\,\,\,\,\, \,\,\,\,\,\,\,\,\,\,
\,\,\,\,\,\,\,\,\,\, \,\,\,\,\,\,\,\,\,\, \,\,\,\,\,\,\,\,\,\,
\,\,\,\,\,\,\,\,\,\,  +\Gamma \left( \epsilon ^{\prime }-\epsilon
-e\mathbf{Er}\right) f\left( \epsilon ^{\prime }\right) \left[
1-f\left( \epsilon \right) \right] \right\}
\label{eq:masterequation}
\end{eqnarray}
by a stationary Fermi distribution.
Equation~(\ref{eq:masterequation}) would follow from our model if
electron correlation (in particular, percolation) effects were not
substantial. In reality, we can expect the results following from
Eq.~(\ref{eq:masterequation}) to be valid only in certain limits
\cite {shklovskiichap2footnote}. For example, in the low field limit
with $E\rightarrow 0$, both methods give $T_{\mathrm{eff}}=T$. At
higher fields the effective temperature grows with the applied
field, which
``overheats" the electrons. At very high fields ($%
E/E_0\gtrsim 0.3$) both methods agree again and give
\begin{equation}
k_{B}T_{\mathrm{eff}}\approx CeEa,\qquad C=0.71\pm 0.02.
\label{eq:strongfieldfitmasterequation}
\end{equation}
(A similar result, but with $C\approx 1.34$, for our definition of
$a$, was obtained by Marianer and Shklovskii
\cite{MarianerShklovskii1992} for a rather different model with an
exponential energy dependence of the density of states $\nu_0$.)
However, at intermediate fields typical of ``high-field"
variable-range hopping ($E_T\ll E\ll E_0$), the master equation
still gives the same result (\ref {eq:strongfieldfitmasterequation})
and hence fails to appreciate that in fact $T_{\mathrm{eff}}$ is
proportional to $E^{2/3}$ \cite{compensation}. In order to explain
this result, let us discuss the statistics of hop lengths
(Figs.~\ref{fig:2dand1dhophistograms} and~\ref{fig:EandThoplength}).

The two- and one-dimensional histograms in
Fig.~\ref{fig:2dand1dhophistograms} show the probability density $P$
of a hop between two sites separated by the vector $\Delta
\mathbf{r}$, and also the density $ P_d$ weighed by the factor
$\left\vert H_{jk}-H_{kj}\right\vert$, where each $H$ is the total
number of hops (during a certain time interval) in the indicated
direction, i.e $jk\equiv j \rightarrow k$. The latter weighing
emphasizes the site pairs $\left( j,k\right)$ contributing
substantially to the net hopping transport, in comparison with
``blinking" pairs which exchange an electron many times before
allowing it to advance along the field. It is clear that at
relatively high temperatures or low fields ($E\ll E_T$) the
non-weighed distribution should be symmetric
(Fig.~\ref{fig:2dand1dhophistograms}a).
Figure~\ref{fig:2dand1dhophistograms}d shows that in this case the
one-dimensional probability density is well approximated by the
Rayleigh distribution, $P\left( r\right)\propto r\exp\left(
-r/a_{\mathrm{eff}} \right)$, with $a_{\mathrm{eff}} \approx a$.
However, the weighed hop distribution is strongly asymmetric even in
the limit $E\rightarrow 0$ (Fig.~\ref{fig:2dand1dhophistograms}b).
This asymmetry is even more evident at low temperatures or high
fields ($E\gg E_T$); in this case the distribution has a sharp
boundary (Fig.~\ref{fig:2dand1dhophistograms}c).
Figure~\ref{fig:2dand1dhophistograms}d shows those cases where the
1D histograms deviate substantially from the distribution predicted
by the master equation - see the first of
Eqs.~(\ref{eq:rmdsprobdens}).

Figure~\ref{fig:EandThoplength} shows the r.m.s$.$ non-weighed
$\left( r_{\mathrm{rms}}\right)$ and direction-weighed $\left(
r_{\mathrm{rmds}}\right)$ hop lengths, defined, respectively, as
\begin{equation}
{r_{\mathrm{rms}}}^2 \equiv {\frac{\sum_{j,k} r_{jk}^2 \left( H_{jk}
+ H_{kj}\right)}{\sum_{j,k} \left( H_{jk}+ H_{kj}\right)}}
\label{eq:rrms}
\end{equation}
and
\begin{equation}
{r_{\mathrm{rmds}}}^2 \equiv \frac{\sum_{j,k} r_{jk}^2 \left\vert
H_{jk}-H_{kj}\right\vert}{\sum_{j,k} \left\vert
H_{jk}-H_{kj}\right\vert}  \label{eq:rrmds}
\end{equation}
(that are of course just the averages of the histograms shown in
Fig.~\ref{fig:2dand1dhophistograms}), as functions of applied
electric field for several values of temperature. At $T\rightarrow
0$, hopping is strictly one-directional (i.e$.$, if $H_{jk}\neq 0$,
then $H_{kj}=0$), so that $r_{\mathrm{rms}}$ and $r_{\mathrm{rmds}}$
are equal. In fact, simulation shows that in this limit
both lengths coincide, at lower fields following the scaling \cite%
{Shklovskii1973}
\begin{equation}
r_{\mathrm{rms}}=r_{\mathrm{rmds}}=Da\left(\frac{E_0}{E}\right)^{1/3},\qquad
E_T\ll E\ll E_0, \label{eq:rrmsrrmdsstrongfield}
\end{equation}
with $D=0.72\pm 0.01$. (We are not aware of any prior results with which this value
could be compared.)

This scaling of $r$ in the variable-range hopping region is
essentially the reason for the scaling of $T_{\mathrm{eff}}$
mentioned above; in fact, the hopping electron gas ``overheating"
may be estimated by equating $k_BT_{\mathrm{eff}}$ to the energy
gain $e\mathbf{E}\mathbf{r}_{\mathbf{rmds}}$, possibly multiplied by
a constant of the order of one. For the effective temperature, this
estimate gives
\begin{equation}
k_BT_{\mathrm{eff}}=const\times
e\mathbf{E}\mathbf{r}_{\mathbf{rmds}}=GeaE_0^{1/3}E^{2/3}=G\frac{\left(
E/E_0\right)^{2/3}}{\nu_0 a^2},
\end{equation}
in accordance with the result shown in
Fig.~\ref{fig:effectivetemperature}. Our Monte Carlo simulations
give $G = 0.60\pm 0.02 $; we are not aware of any previous results
with which this number could be compared.

At higher fields $\left(E\gtrsim 0.1E_0\right)$ the hop lengths
start to decrease slower, approaching a few localization lengths $a$
(Fig.~\ref{fig:EandThoplength}a). In this (``ultra-high-field")
region, the energy range $\sim eEa$ for tunneling at distances of a
few $a$ is so high that there are always some accessible empty sites
within this range, so that long hops, so dominant at variable-range
hopping, do not contribute much into conduction.

At finite temperatures, the most curious result is a non-monotonic
dependence of the r.m.s$.$ hopping length on the applied field - see
Fig.~\ref {fig:EandThoplength}a. At $E\rightarrow 0$,
$r_{\mathrm{rms}}$ has to be field-independent, and there is no
scale for it besides $a$. (As evident as it may seem, this fact is
sometimes missed in popular descriptions of hopping.) In order to
make a crude estimate of $r_{\mathrm{rms}}$ in this limit, one can
use the master equation (\ref{eq:masterequation}). In this approach,
at thermal equilibrium (i.e$.$ at $\Gamma$ independent of $r$), the
hop length probability density $P\left( r\right)$ can be found as
\begin{eqnarray}
P\left( r\right) & = & 2\pi r \int d\epsilon \int d\epsilon^{\prime
}\exp \left( \frac{-r}{a}\right) \left\{\Gamma \left( \epsilon
-\epsilon ^{\prime }\right) f\left( \epsilon \right) \left[
1-f\left( \epsilon ^{\prime }\right) \right] \right.
\nonumber \\
&& \left. \,\,\,\,\,\,\,\,\,\, \,\,\,\,\,\,\,\,\,\,
\,\,\,\,\,\,\,\,\,\, \,\,\,\,\,\,\,\,\,\, \,\,\,\,\,\,\,\,\,\,
\,\,\,\,\,\,\,\,\,\, \,\,\,\,\,\,\,\,\,\, \,\, +\Gamma\left(
\epsilon^{\prime }-\epsilon \right) f\left( \epsilon ^{\prime
}\right) \left[ 1-f\left( \epsilon \right) \right] \right\}
\nonumber
\\
& \propto & r\exp \left( \frac{-r}{a}\right), \label{eq:rmsprobdens}
\end{eqnarray}
in a good agreement with the results shown in
Fig.~\ref{fig:2dand1dhophistograms}d for this case. From
Eq.~(\ref{eq:rmsprobdens}), we get
\begin{equation}
r_{\mathrm{rms}}\equiv \left\langle r^2\right\rangle^{1/2}=\left[
\frac{\int P\left( r\right) r^2 dr}{\int P\left( r\right)dr}
\right]^{1/2}=\sqrt{6} a\approx 2.45a,
\end{equation}
in a good agreement with numerical data shown in
Fig.~\ref{fig:EandThoplength}.

A similar calculation for $r_{\mathrm{rmds}}$ may be obtained by
expanding the tunneling rate $\Gamma(\epsilon-\epsilon ^{\prime
}+e\mathbf{Er})$ in small electric field as
$\Gamma(\epsilon-\epsilon ^{\prime })+e\mathbf{Er} \Gamma^{\prime
}(\epsilon-\epsilon ^{\prime })$. The result is
\begin{eqnarray}
P_d\left( r\right) & = & eEr^2 \int d\phi \left\vert cos\phi
\right\vert \int d\epsilon \int d\epsilon^{\prime }\exp \left(
\frac{-r}{a}\right)
\nonumber \\
&& \times \left\{\Gamma^{\prime } \left( \epsilon^{\prime
}-\epsilon\right) f\left( \epsilon^{\prime } \right) \left[
1-f\left( \epsilon \right) \right] +\Gamma^{\prime } \left( \epsilon
-\epsilon^{\prime } \right) f\left( \epsilon \right) \left[
1-f\left( \epsilon^{\prime } \right) \right]
\right\},  \nonumber \\
P_d\left( r\right)& \propto & r^2\exp\left(\frac{-r}{a}\right),  \nonumber \\
r_{\mathrm{rmds}} & \equiv & \left[ \frac{\int P_d\left( r\right)
r^2 dr}{\int P_d\left( r\right)dr} \right]^{1/2}=\sqrt{12} a\approx
3.46a. \label{eq:rmdsprobdens}
\end{eqnarray}
The Monte Carlo data (Fig.~\ref{fig:EandThoplength}), however,
differ from this result \cite{masteqvalidityfootnote}, showing that
at $E\rightarrow 0$, $r_{\mathrm{rmds}}$ obeys the Mott law
\cite{MottBook,ShklovskiiBook,EfrosPollackCollection}
\begin{equation}
r_{\mathrm{rmds}}=Ha\left(\frac{T_0}{T}\right)^{1/3}+Ia,\qquad T\ll
T_0, \label{eq:highTrrmdsfit}
\end{equation}
with the best-fit values $H=0.52\pm 0.05$ and $I=2.0\pm 0.1$. In
contrast, the function $r_{\mathrm{rms}}\left(T\right)$ is rather
far from Eq.~(\ref{eq:highTrrmdsfit}), because the Mott law refers
to long hops responsible for transport (with the average
approximately corresponding to $r_{\mathrm{rmds}}$), while
$r_{\mathrm{rms}}$ reflects the statistics of all hops.

\section{\label{sec:level1}Shot Noise}

The current noise simulation has been limited to the case of zero
temperature for two reasons. First, in the opposite limit $\left(E_T
\gg E \right)$ current noise obeys the fluctuation-dissipation
theorem; as a result, its low-frequency intensity can be found from
$\sigma(T)$, and hence does not provide any new information. Second,
the calculation of the spectral density $S_I \left( f \right)$ of
current fluctuations with acceptable accuracy requires much larger
statistical ensemble of random samples than that of the average
current, which means it becomes increasingly difficult to extend it
to finite temperatures even with the advanced averaging algorithm
and substantial supercomputer resources used in this work.

\begin{figure}
\begin{center}
\includegraphics[height=3.4in]{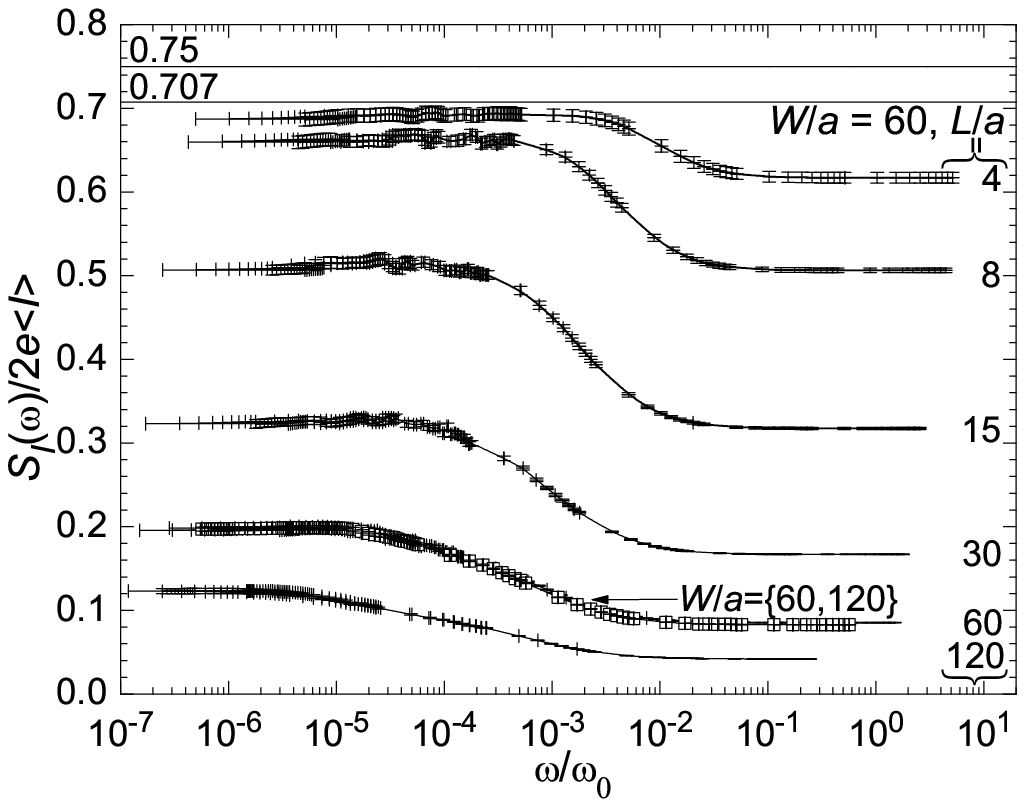}
\end{center}
\caption{Spectral density $S_I$ of current fluctuations normalized
to the Schottky noise $2e\left\langle I\right\rangle$ as a function
of observation frequency $\omega$, measured in units of $\omega_{0}
\equiv g/\hbar \nu_0 a^2$, for several values of sample length $L$
for $T=0$ and $E/E_0=8.75\times10^{-3}$. Small points show results
for $W/a=60$, while open squares are for $W/a = 120$ (at $L/a=60$).
Horizontal lines correspond to the Fano factor for hopping through
short samples with one and two localized sites
\cite{1SiteNazarovStruben1996,2SiteKinkhabwalaKorotkov2000}. Curves
are only guides for the eye.} \label{fig:specdensfrequency}
\end{figure}

Figure~\ref{fig:specdensfrequency} shows a typical dependence of the
current noise spectral density $S_I$, normalized to the Schottky
value $2e \left\langle I \right\rangle$, on the observation
frequency $\omega \equiv 2 \pi f$. One can see a crossover from a
low-frequency plateau to another plateau at high frequencies. As the
sample length grows, the crossover becomes extended, i.e.~features a
broad intermediate range $\omega_l \leq \omega \leq \omega_h$, just
like in 1D systems with next-site hopping
\cite{1DKorotkovLikharev2000}. The position of the high-frequency
end $\omega_h$ of this region can be estimated in the following way.

In all single-electron tunneling systems, the high-frequency plateau
is reached at frequency $\omega_h$ close to the rate $\Gamma$ of the
fastest electron hops affecting the total current
\cite{1DKorotkovLikharev2000,2DSverdlovKorotkovLikharev2001,KorAveLikVas,Kor-94}.
(For example, in systems described by the ``orthodox" theory of
single-electron tunneling, $\omega_h \approx \Gamma_{\mathrm{max}}
\approx \Delta U_{\mathrm{max}}/e^2R \approx 1/RC$, where $R$ and
$C$ are, respectively, resistance and capacitance of a single
junction \cite{KorAveLikVas,Kor-94}.) In our current case, this
means that $\omega_h \sim (\Gamma_{jk})_{\mathrm{max}} \sim
(g/\hbar)[\Delta U_{jk} \exp{(-r_{jk}/a)}]_{\mathrm{max}}$. For this
estimate, $\Delta U_{jk}$ can be taken as $k_BT_{\mathrm{eff}}$ from
Fig.~\ref{fig:effectivetemperature}, while according to the
histograms shown in Fig.~\ref{fig:2dand1dhophistograms}(d), the
length of shortest hops, still giving a noticeable contribution to
the current, can be estimated as $\sim r_{\mathrm{rmds}}/2$. For the
case shown in Fig.~\ref{fig:specdensfrequency} ($E/E_0 = 8.75 \times
10^{-3}$), these estimates yield $\Delta U_{jk}/k_B T_0 \approx 2
\times 10^{-2}$, $(r_{jk})_{\mathrm{min}}/a \approx 2.5$, giving
finally $\omega_h/\omega_0 \sim 1.5 \times 10^{-3}$, in a very
reasonable agreement with numerical results shown in
Fig.~\ref{fig:specdensfrequency}. (Note that this simple estimate,
giving a length-independent value for $\omega_h$, is only valid for
relatively long and broad samples.)

At $\omega \rightarrow 0$, a crossover to $1/f$ noise might be
expected, because the discussion of this effect in some earlier
publications \cite{Shklovskii1980_1981,Shklovskii2003} was
apparently independent of the Coulomb interaction between hopping
electrons. However, within the accuracy of our simulations, we could
not find any trace of $1/f$-type noise for any parameters we have
explored. This fact may not be very surprising, because all the
discussions of the $1/f$ noise we are aware of require the presence
of thermal fluctuations which are absent in our case ($T=0$).

Since the low-frequency spectral density is flat, at $T = 0$ it may
be considered as shot noise \cite{specdenscurrentfootnote} and
characterized by the Fano factor \cite{BlanterButtiker2000}
\begin{equation}
F\equiv \frac{S_I\left( f \rightarrow 0\right)}{2e\left\langle
I\right\rangle}. \label{eq:fanofactor}
\end{equation}
Similarly, in order to
characterize the flat high-frequency spectral density, we may use
the parameter
\begin{equation}
F_{ \infty } \equiv \frac{S_I\left( f \rightarrow  \infty
\right)}{2e\left\langle I\right\rangle}.
\label{eq:inffreqfanofactor}
\end{equation}

\begin{figure}
\begin{center}
\includegraphics[height=5.4in]{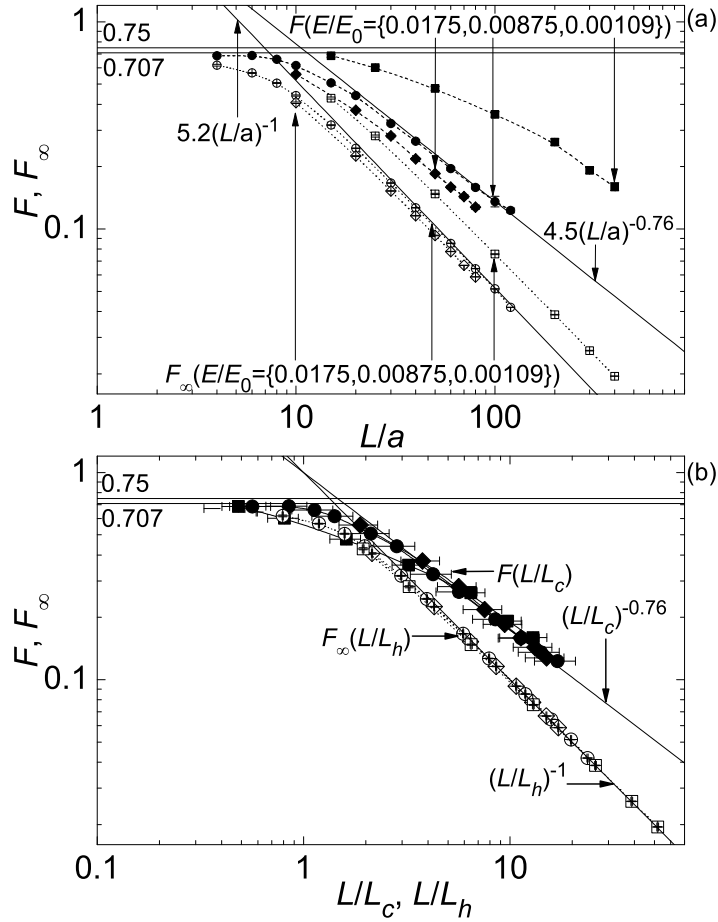}
\end{center}
\caption{Average Fano factor $F$ and its high-frequency counterpart
$F_{\infty}$ as functions of sample length $L$ normalized to: (a)
the localization length $a$, and (b) the scaling lengths $L_c$ (for
$F$) and $L_h$ (for $F_{\infty}$) (see Fig.~\ref{fig:clusterlength}
below), for several values of applied field at $T=0$ and $W\gg L_c$.
Horizontal lines correspond to the average Fano factor for hopping
via one and two localized sites
\cite{1SiteNazarovStruben1996,2SiteKinkhabwalaKorotkov2000}.
Straight lines are the best fits to the data, while dashed curves
are only guides for the eye.} \label{fig:fanovslength}
\end{figure}

Figure~\ref{fig:fanovslength}(a) shows the average Fano factors
$F$ and $F_{\infty}$ as a function of $L$ for several values of
the applied electric field, while Fig.~\ref{fig:fanovslength}(b)
shows that the same data can be collapsed on universal curves by
the introduction of certain length scales: $L_c$ for $F$ and $L_h$
for $F_{\infty}$.

\begin{figure}
\begin{center}
\includegraphics[height=3.4in]{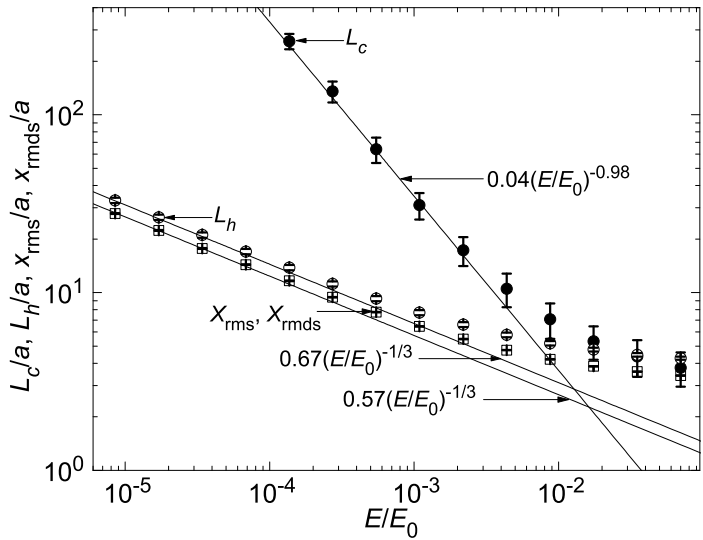}
\end{center}
\caption{The values of $L_h$ and $L_c$ giving the best fitting of
shot noise results by Eqs.~(\ref{eq:infinitefrequencyfanolength})
and~(\ref{eq:fanolength}) as functions of applied electric field $E$
(open and solid circles, respectively). Open squares show the simple
and direction-weighed average hop length along the applied field
direction, defined similarly to Eqs.~(\ref{eq:rrms})
and~(\ref{eq:rrmds}). Straight lines are the best fits to the data.}
\label{fig:clusterlength}
\end{figure}

For the high frequency case,
\begin{equation}
F_{\infty} = \left( \frac{L_h}{L}\right)^{\beta},\qquad L\gg L_h,
\label{eq:infinitefrequencyfanolength}
\end{equation}
where, within the accuracy of our calculations, $\beta = 1$. Such
dependence could be expected, because the high-frequency noise at
hopping can be interpreted as a result of the ``capacitive division"
of the discrete increments of externally-measured charge jumps
resulting from single-electron hops through the system
\cite{Av-Likh}. When applied to uniform (ordered) systems, these
arguments always give the result $F_{\infty} \approx 1/N_h$ with
$N_h=L/d$ being the number of electron hops ($d$ the hopping length
along the current flow) necessary to pass an electron through the
system, regardless of hop rates
\cite{BlanterButtiker2000,MatsuokaLikharev1998,1DKorotkovLikharev2000}.
For $T \rightarrow 0$ in the case of disordered conductors, $L_h$ in
Eq.~(\ref{eq:infinitefrequencyfanolength}) may then be interpreted
as the average hop length along current flow. This interpretation
turns out to be correct. Indeed, Fig.~\ref{fig:clusterlength} shows
that the parameter $L_h$ obtained from
Eq.~(\ref{eq:infinitefrequencyfanolength}) scales with the electric
field in a manner similar to $x_{\mathrm{rmds}}$, especially at low
fields, where it follows the variable-range hopping dependence of
Eq.~(\ref{eq:rrmsrrmdsstrongfield}).

The low-frequency value $F$, in the limit ($L \ll L_c$), is weakly
dependent on length and approaches $F \approx 0.7$, which not
surprisingly is consistent with the prior results for hopping via 1
intermediate site \cite{1SiteNazarovStruben1996} ($F=0.75$) and 2
such sites \cite{2SiteKinkhabwalaKorotkov2000} ($F=0.707$). The
results for long samples are much more interesting. We have found
that they may be reasonably well fit with a universal dependence
\begin{equation}
F = \left( \frac{L_c}{L}\right)^{\alpha},\qquad L\gg L_c.
\label{eq:fanolength}
\end{equation}
Here $\alpha$ is a numerical exponent; in the current study we
could establish that
\begin{equation}
\alpha=0.76\pm 0.08.
\end{equation}
Equation~(\ref{eq:fanolength}) and the value for $\alpha$ are
compatible with our previous results
\cite{2DSverdlovKorotkovLikharev2001} $\alpha=0.85\pm 0.07$ (for the
same model, but just one particular value of $E$) and
$\alpha=0.85\pm 0.02$ (for nearest neighbor hopping on uniform
slanted lattices).

Figure~\ref{fig:clusterlength} shows the fitting parameter $L_c$ as
a function of electric field $E$. In the variable-range hopping
region, it may be fitted with the following law,
\begin{equation}
L_c=Ja\left( \frac{E_0}{E} \right)^{\mu},\qquad J=0.04\pm 0.01, \
\mu=0.98\pm 0.08.  \label{eq:clusterlengthstrongfield}
\end{equation}
This law may be compared with the result of the following arguments.
According to the arguments given in
Ref.~\cite{2DSverdlovKorotkovLikharev2001}, parameter $L_c$ may be
interpreted as the average percolation cluster length (up to a
constant of the order of $1$). The theory of directed percolation
\cite{StaufferAharonyBook,Obukhov1980,Essametal1986} gives the
following scaling:
\begin{equation}
L_c\propto \left\langle x\right\rangle \left( \frac{x_c}{\left\vert
\left\langle x\right\rangle-x_c\right\vert}
\right)^{\delta_{\parallel}}.
\end{equation}
Here $\left\langle x\right\rangle$ is the r.m.d.s$.$ hop length
along the field direction, while $x_c$ is its critical value.
According to Ref.~\cite{Essametal1986}, the critical index
$\delta_{\parallel}$ should be close to $1.73$. Due to the
exponential nature of the percolation, $\left\vert\left\langle
x\right\rangle - x_c\right\vert \sim a$, while $\left\langle
x\right\rangle$ should follow a field scaling similar to that given
by Eq.~(\ref{eq:rrmsrrmdsstrongfield}). (Square points in
Fig.~\ref{fig:clusterlength} show that this is true for our
simulation results as well.) Thus for sufficiently large
$\left\langle x\right\rangle$ we arrive at
Eq.~(\ref{eq:clusterlengthstrongfield}) with $\mu=\frac{1}{3}\left(
1+\delta_{\parallel}\right)\approx 0.911$. Equation (24) shows that
this value is quite compatible with our numerical result, thus
confirming the interpretation of $L_c$ as the average percolation
cluster length.

Note that in the variable-range hopping regime, $L_c$ has a
different field dependence and is much larger than the average hop
length. However, as the applied electric field approaches $E_0$,
both lengths become comparable with each other and with the
localization radius $a$.

\section{\label{sec:level1}Discussion}

To summarize, our results for average conductivity and hop
statistics are in agreement with the well-known semi-quantitative
picture of hopping, including the usual variable-range hopping at
low fields $\left(E\ll E_T\right)$ and ``high-field" variable-range
hopping at $E_T\ll E\ll E_0$. However, our supercomputer-based
simulation has allowed, for the first time, a high-precision
quantitative characterization of hopping, for a particular but very
natural model. Moreover, our model also describes the ``ultra-high
field" region $ \left(E\sim E_0\right)$ where the variable-range
hopping picture is no longer valid, since from most localized sites
an electron can hop, with comparable probability, to several close
sites. (In the last region, there are no clearly defined percolation
clusters; rather, electrons follow a large number of interwoven
trajectories.)

Our simulations of shot noise at 2D hopping have confirmed our
earlier hypothesis \cite{2DSverdlovKorotkovLikharev2001} that in the
absence of substantial Coulomb interaction, in sufficiently large
samples $\left( L,W\gg L_c\right)$ the Fano factor $F$ scales
approximately proportionally to $1/L$ - see
Eq.~(\ref{eq:fanolength}). Other confirmations of this hypothesis
have come from recent experiments with lateral transport in SiGe
quantum wells  \cite{Kuznetsovetal2000} and GaAs MESFET channels
\cite{Roshkoetal2002,Savchenkoetal2004}. Unfortunately, these
experiments are not precise enough to distinguish the small
difference between the exponent $\alpha$ in
Eq.~(\ref{eq:fanolength}) and unity.

The hypothesis that $\alpha$ is in fact equal to 1 for sufficiently
long samples seems appealing, because it would mean the simple
addition of mutually-independent noise voltages generated by sample
sections connected in series. On the contrary, a deviation of
$\alpha$ from unity would mean that some dynamic correlations of
electron motion persist even at $L \gg L_c$. For 1D hopping this
fact is well established: in at least one exactly solvable model
(the ``asymmetric single exclusion process", or ASEP \cite{Derrida})
the dynamic correlations may change $\alpha$ from 1 to 0.5. However,
for 2D conductors the fact that the correlation length (if any) may
be substantially larger than the percolation cluster length comes as
a surprise \cite{reliable}.

From the point of view of possible applications in single-electron
devices \cite{LikharevIEEE1999}, the fact that $F$ may be suppressed
to values much less than unity is generally encouraging, since it
enables the use of circuit components with quasi-continuous charge
transport. However, in order to achieve the high-quality
quasi-continuous transfer (say, $F\lesssim 0.1$), the sample length
$L$ has to be at least an order of magnitude longer than the
percolation cluster length scale $L_c$. On the other hand, $L_c$
itself, especially in the most interesting case of low applied
fields, is substantially longer than the localization length $a$
(Fig.~\ref{fig:clusterlength}), which is of the order of a few
nanometers in most prospective materials, e.g., amorphous silicon.
Hence, it may be hard to implement sub-electron transport in
conductors substantially shorter than $\sim$100 nm. This size is
quite acceptable for experiments at low temperatures (say, below 1
K), but is too large for the most important case of room-temperature
single-electron circuits \cite{LikharevIEEE1999}, because of large
stray capacitance of the hopping conductor, which effectively sums
up with the capacitance of the island to be serviced.

\ack{Fruitful discussions with A. Efros, V. Kozub, M. Pollak, B.
Shklovskii and D. Tsigankov are gratefully acknowledged. The work
was supported in part by the Engineering Physics Program of the
Office of Basic Energy Sciences at the U.S. Department of Energy. We
also acknowledge the use of the following supercomputer resources:
our group's's cluster $Njal$ (purchase and installation funded by
DoD's DURIP program via AFOSR), Oak Ridge National Laboratory's IBM
SP computer $Eagle $ (funded by the Department of Energy's Office of
Science and Energy Efficiency program), and also IBM SP system
$Tempest$ at Maui High Performance Computing Center and IBM SP
system $Habu$ at NAVO Shared Resource Center. Computer time on the
two last machines has been granted by DOD's High Performance
Computing Modernization Program.}

\end{document}